\documentclass{elsart}

\usepackage{graphicx}
\usepackage{amssymb}

\begin{document}

\begin{frontmatter}



\title{Magnetic properties and thermal expansion of YbGa$_{x}$Ge$_{2-x}$ ($0.9 \leq x \leq 1.5$)}

\author{N. Tsujii\corauthref{cor1},}
\ead{TSUJII.Naohito@nims.go.jp}
\corauth[cor1]{Corresponding author, Tel.+81-29-859-2817, Fax +81-29-859-2801}
\author{T. Furubayashi, H. Kitazawa, G. Kido}
\address{National Institute for Materials Science, Sengen 1-2-1, Tsukuba 305-0047, Japan}

\begin{abstract}
The composition dependence of the Yb valence and of the thermal expansion have been studied
in the YbGa$_x$Ge$_{2-x}$ system.
X-ray diffraction reveals that single-phased samples isostructural to YbGaGe
are obtained in the range of $0.9 \leq x \leq 1.5$.
The magnetic susceptibility shows that the systems are almost nonmagnetic below room temperature, 
indicating a stable Yb$^{2+}$ state for the $x$ range $0.9 \leq x \leq 1.5$.
The lattice constants as well as the lattice volume of these systems are found to decrease
monotonically with decreasing temperature, suggesting the absence of zero thermal-expansion
previously reported for YbGaGe.
\end{abstract}

\end{frontmatter}

\section{Introduction}
Yb-based intermetallic compounds are of interest due to their anomalous physical 
properties concerning to valence fluctuation and/or heavy-fermion behavior~\cite{Bauer,Laewenhaupt}.
Yb ion can have two different valence state: divalent(Yb$^{2+}$) and trivalent(Yb$^{3+}$) states.
The former has a closed 4$f$ shell and thereby is nonmagnetic, whereas the latter
has one hole in the 4$f$ shell, leading to a total angular momentum of $J$ = 7/2.
Valence-fluctuating Yb-compounds are characterized by nearly trivalent
Yb ions at temperatures sufficiently higher than the valence-fluctuation temperature, $T_{\rm VF}$.
With decreasing temperature, valence admixture with the divalent state occurs,
resulting in a nonmagnetic ground state with a large density of states.
Since divalent Yb has a larger ionic radius than trivalent Yb,
lattice expansion of Yb-based compounds can occur on cooling.
The most prominent example may be YbInCu$_4$, 
which shows a first-order valence-transition at $T$ = 40 K~\cite{Felner},
at which the lattice abruptly expands more than 0.1\%{} on cooling~\cite{Felner2}.

If such volume expansion occurs moderately to cancel the normal lattice shrinkage with decreasing temperature,
the total volume change can become zero.
Such zero thermal-expansion has recently been reported by Salvador et al. in the intermetallic compound
YbGaGe~\cite{Salvador}.
It has been reported that the volume change is almost negligible in a wide temperature
range from 300 K to 100 K~\cite{Salvador}.
The authors have also presented magnetic susceptibility data that suggest a Yb valence changes
from Yb$^{3+}$ at high temperatures ($\gtrsim$150 K) to nearly divalent state at low temperatures~\cite{Salvador}.
From these facts, they have concluded that the zero thermal-expansion is due to the valence change of Yb.

Lately, however, controversial results have been reported for YbGaGe by several authors
~\cite{Margadonna,Muro,Bobev}.
Margadonna et al. have reported the absence of zero thermal-expansion.
They have reported that the lattice volume shrinks similarly to normal substances below 700 K~\cite{Margadonna}.
Instead, these authors have found a sudden volume change at 5 K without symmetry change, possibly due
to a valence transition of Yb~\cite{Margadonna}.
On the other hand, Muro et al. and Bobev et al. have shown that the magnetic susceptibility of YbGaGe
corresponds to an almost nonmagnetic state, suggesting the Yb$^{2+}$ state in whole the temperature range~\cite{Muro,Bobev}.
These results are indicative of large sensitivity of magnetic and thermal properties of YbGaGe
to the chemical composition or stoichiometry.

To clarify these discrepancies, we have studied the composition dependence of the physical properties 
of the YbGaGe system.
In this paper, we report the solubility range of YbGa$_x$Ge$_{2-x}$ system,
and their magnetic susceptibility and the thermal expansion.

\section{Experimental}
Polycrystalline samples were prepared by argon arc melting from pure metals.
To compensate the loss of volatile Yb, a slightly richer amount of Yb ($3\sim5$\%{}) was charged.
The samples were subsequently annealed in evacuated silica tubes at 750$^{\circ}$C for 4 days.
For samples with Ge-rich compositions, a higher annealing temperature (850$^{\circ}$C) for 1 day
was also examined.
EDS analysis revealed that the samples thus prepared were almost of ideal chemical composition.
Structure of samples were studied by powder X-ray diffraction at room temperature
using Cu K$_{\alpha}$ radiation with a RIGAKU diffractmeter.
The magnetic susceptibility was measured with a SQUID magnetometer (Quantum Design, MPMS5S)
at 1000 Oe.
The temperature dependence of the lattice constants was investigated 
by means of powder X-ray diffraction using another RIGAKU diffractmeter.

\section{Results and discussion}
\begin{figure}[t]
 \begin{center}
 \includegraphics[width=10cm]{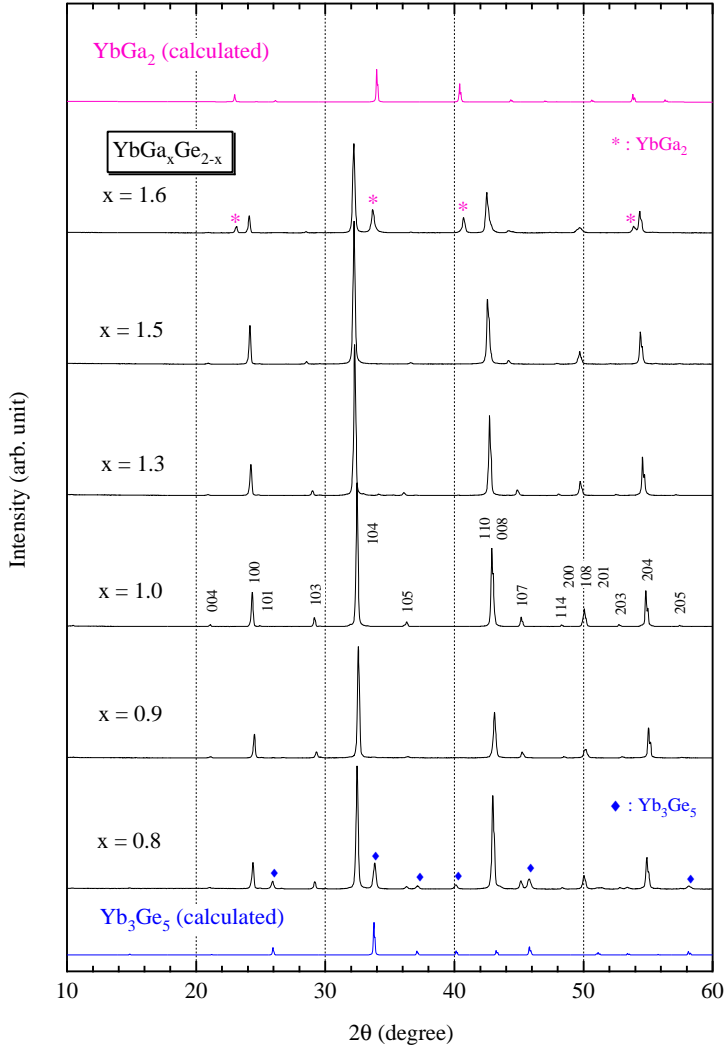}
 \caption{
 Powder X-ray diffraction patterns of YbGa$_x$Ge$_{2-x}$.
 Patterns of YbGa$_2$ and Yb$_3$Ge$_5$ are calculated ones.
 Peaks marked by $\ast$ are due to the YbGa$_2$ phase,
 and those marked by $\blacklozenge$ are due to Yb$_3$Ge$_5$.
 }
 \end{center}
\end{figure}
At first, we have studied the stoichiometry for Yb$_{1\pm{}\delta}$GaGe.
SEM and EDS analysis have shown that the solubility range is within 3\%,
indicating that the chemical composition ratio of Yb:(Ga,Ge) is almost 1 : 2.
\begin{figure}[tp]
 \begin{center}
 \includegraphics[width=8cm]{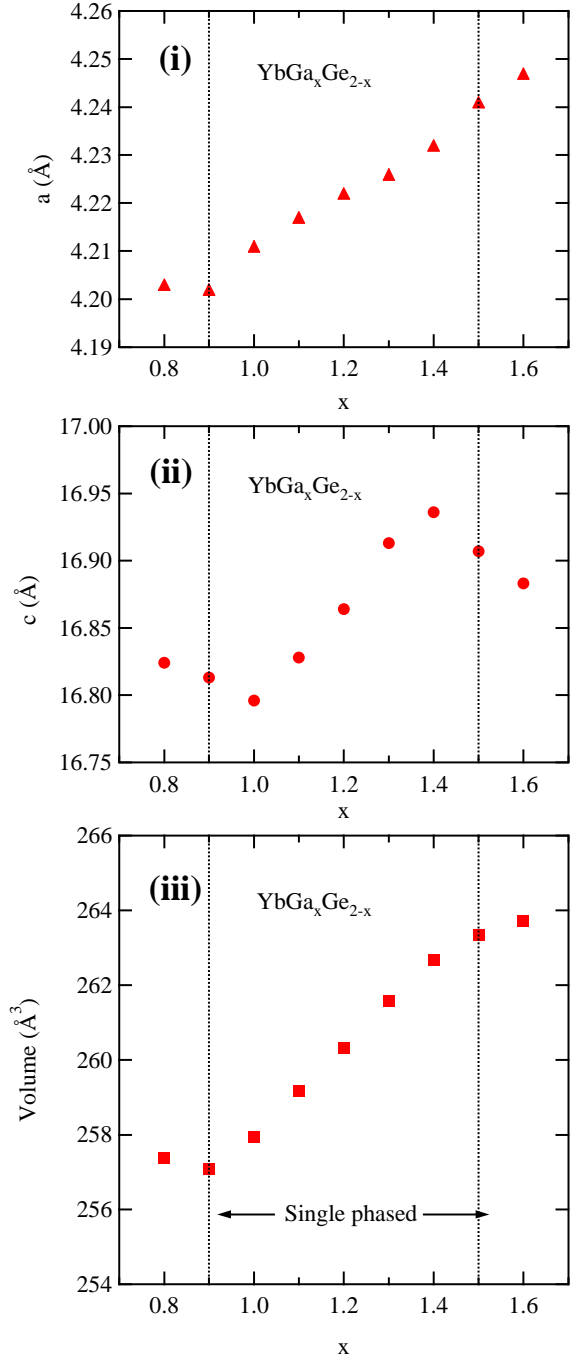}
 \caption{
 Lattice constants $a$, $c$, and unit cell volume of YbGa$_x$Ge$_{2-x}$ at room temperature.
 Dotted lines indicate the phase boundaries, inside which
 single-phased samples are obtained.
 }
 \end{center}
\end{figure}

Next, we have studied the system YbGa$_x$Ge$_{2-x}$.
In Fig.1, powder X-ray diffraction patterns of YbGa$_x$Ge$_{2-x}$ are shown.
The pattern of $x$=1.0, i.e., YbGaGe, well agrees with a calculation (not shown)
based on the structure parameter reported in ref~\cite{Salvador}.
It is seen in the figure that the YbGa$_x$Ge$_{2-x}$ compounds have the same crystal structure
up to $x$ = 1.5, i.e., YbGa$_{1.5}$Ge$_{0.5}$.
For $x > 1.5$, additional peaks are found to develop, which are attributed to the YbGa$_2$ phase
with CaIn$_2$-type structure~\cite{Schwarz}, as is seen in the figure,
where the peaks from the YbGa$_2$ phase are marked by asterisks.

For $x < 1.0,$ the solubility range has been found to be narrow.
In Fig.1, the patterns for $x$=0.9 and 0.8 are shown.
One can see that peaks due to the Yb$_3$Ge$_5$ phase appear for $x \leq 0.8$.
These data are those of the samples annealed at 850$^{\circ}$C and subsequently quenched.
For samples annealed at 750$^{\circ}$C, the Yb$_3$Ge$_5$ phase is seen even for $x$ = 0.9,
suggesting that the solubility range is narrower at low temperatures.
From these results, it is concluded that the single-phase region of YbGa$_x$Ge$_{2-x}$
is $0.9 \leq x \leq 1.5$.

\begin{figure}[tbp]
 \begin{center}
 \includegraphics[width=15cm]{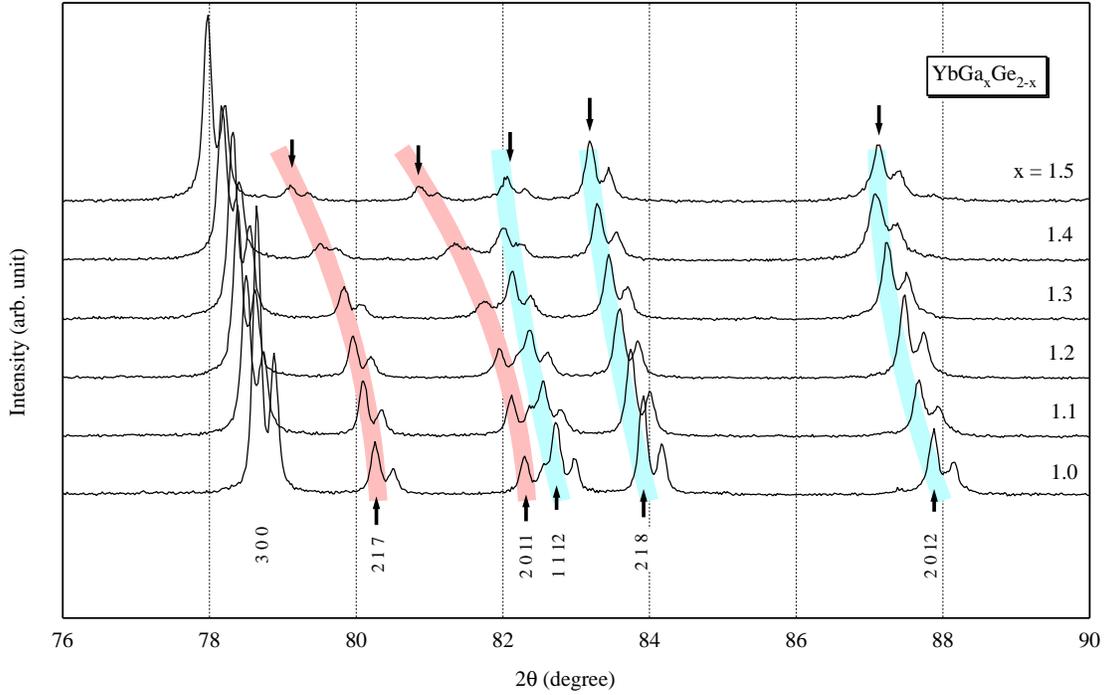}
 \caption{
 X-ray diffraction patterns of YbGa$_x$Ge$_{2-x}$ at high angles.
 Bold lines are guide to eye.
 }
 \end{center}
\end{figure}
In Fig.2, the lattice constants of YbGa$_x$Ge$_{2-x}$ are shown.
These lattice constants were determined by a least square method 
using the peak position of the (3 0 0), (1 1 12), (2 1 8), and (2 0 12) reflections.
It is found that with increasing Ga concentration $x$, $a$ increases monotonically.
On the other hand, $c$ does not show a monotonic composition dependence.
The origin of this is not yet clear.
Nevertheless, the unit cell volume varies monotonically with $x$ in the single-phase region
($0.9\leq x \leq 1.5$).
We have found that the standard deviation for the estimation of lattice constants
becomes extremely large for $x \geq 1.4$ when we use the ($h k l$) peaks with odd $l$ numbers,
such as (2 1 7) and (2 0 11).
This is probably due to stacking faults along the $c$-direction for Ga-rich compositions,
as is shown below.

In Fig.3, X-ray diffraction patterns of YbGa$_x$Ge$_{2-x}$ at high angles are shown.
One can see that the (3 0 0) peak monotonically shifts to lower angles with increasing $x$,
indicating the monotonic increase of $a$.
On the other hand, other reflections do not show such simple $x$-variations.
For the (2 1 7) and (2 0 11) reflections, the peaks shift to lower angles with increasing $x$.
Especially for $x \geq 1.4$, the peaks shift rapidly.
In contrast, the (2 1 8) and (2 0 12) reflections do not shift rapidly to lower angles, 
or even shift to higher angles for $x \geq 1.4$.
It is also noticeable that the intensities of the peaks with odd $l$ show
a remarkable decrease for $x \geq 1.4$, whereas those with even $l$ are
almost unchanged.
These $x$ variations cannot be explained by an $x$-dependence of the lattice parameters.
It should be noted that for YbGaGe, the Ga-Ge layers are stacked along the $c$ direction 
with the sequence $ABCDABCD\cdot\cdot$~\cite{Salvador}.
On the other hand, for YbGa$_2$, the Ga layers stack as $ABAB\cdot\cdot$~\cite{Schwarz}.
We speculate for YbGa$_x$Ge$_{2-x}$ with large $x$, that the stacking of $ABAB$ is inserted
in the regular $ABCD$ stacking.
For the case of complete $ABAB$ stacking, ($hkl$) reflection with odd $l$ should vanish.
Thereby the insertion of $ABAB$ in the regular $ABCD$ stacking can lead to an incommensurate modulation
of the $d$-value of these plains, giving rise to an $x$-dependence of the peak positions
different from that expected from the lattice constants.

\begin{figure}[t]
 \begin{center}
 \includegraphics[width=10cm]{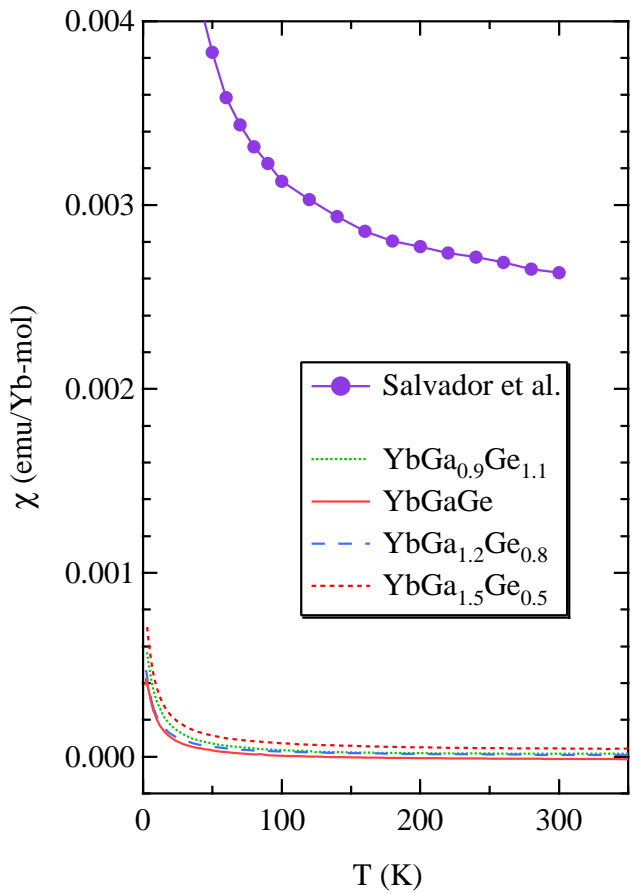}
 \caption{
 Magnetic susceptibility of YbGa$_x$Ge$_{2-x}$ measured at $H$ = 1000 Oe.
 Data from ref.~\cite{Salvador} are also shown.
 }
 \end{center}
\end{figure}
Having determined the single phase region to be $0.9 \leq x \leq 1.5$,
we show the physical properties of the YbGa$_x$Ge$_{2-x}$ system.
In Fig.4, the magnetic susceptibility $\chi$ of YbGa$_x$Ge$_{2-x}$ is shown together with that 
reported by Salvador et al. in ref.~\cite{Salvador}.
In a good contrast to the data of ref.~\cite{Salvador}, our data show
that the susceptibility of YbGa$_x$Ge$_{2-x}$ is extremely small in whole the range of $x$.
We have fitted the data between 150 and 300 K using a Curie-Weiss function:
$\chi(T) = C/(T-\theta) + \chi_0$,
where $C$, $\theta$, and $\chi_0$ are the Curie constant, Weiss temperature, and
temperature-independent susceptibility, respectively.
Here $C$ is written as $C = \alpha Np_{\rm eff}^2/3k_{\rm B}$,
where $\alpha$ is the concentration of Yb$^{3+}$ per formula unit,
$N$ the Avogadro number, $p_{\rm eff}$ the effective magnetic moment,
and $k_{\rm B}$ the Boltzmann constant.
$p_{\rm eff}$ should be 4.54${\rm \mu_B}$ for the case of Yb$^{3+}$.
The results of fitting have yielded a value of $C < 0.003$, which corresponds
to $\alpha \lesssim 0.1\%$.
This indicates that most of the Yb ions are nonmagnetic, i.e., in the Yb$^{2+}$ state.
This result is consistent with the results of X-ray photoemission spectroscopy measurement
on polycrystalline samples~\cite{Yokoya}.
The low temperature increase in $\chi$ is then attributed to a small amount of impurities.
Since the Yb$^{2+}$ state is consistently dominant for all the $x$ measured,
we conclude that the valence of the Yb ions in YbGa$_x$Ge$_{2-x}$ is 
divalent, independent of the chemical composition.

In addition, no magnetic transition was observed down to 3 K.
In several samples, a very tiny transition was observed at 2.3 K,
and this is attributed to the antiferromagnetic transition in a trace of Yb$_2$O$_3$,
of which the N\'{e}el temperature is $T_{\rm N}$ = 2.2 K~\cite{Klaasse}.
This contrasts with the valence transition around 5 K suggested in ref.~\cite{Margadonna}.
\begin{figure}[t]
 \begin{center}
 \includegraphics[width=15cm]{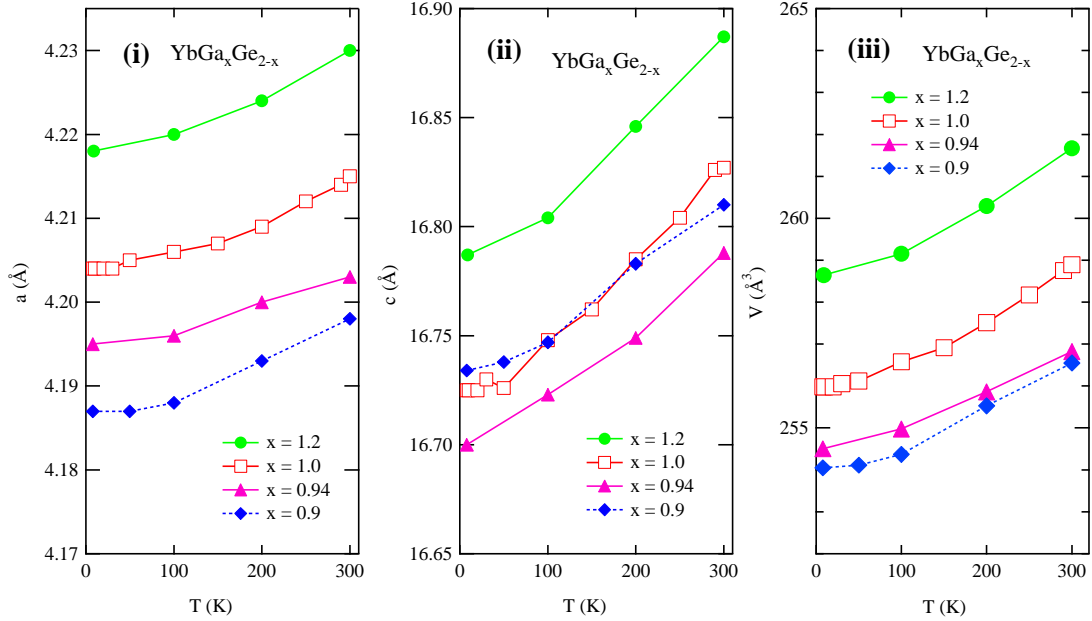}
 \caption{
 Temperature dependence of the lattice constant $a$, $c$, and the lattice volume $V$
 of YbGa$_x$Ge$_{2-x}$.
 }
 \end{center}
\end{figure}

Since the ionic radius of Yb$^{2+}$ is larger than that of Yb$^{3+}$, 
a volume expansion with decreasing temperatures or zero thermal-expansion should not occur
for these systems, as far as the valence-change mechanism is concerned.
To confirm this expectation, we have studied the thermal expansion of
YbGa$_x$Ge$_{2-x}$, of which results are shown in Fig. 5.
For YbGaGe, both $a$ and $c$ decrease monotonically with decreasing temperatures
as in normal substances.
As a result, the lattice volume also shows a monotonical decrease on cooling.
This result for $x$ = 1.0 is consistent with that reported for YbGaGe 
by several authors~\cite{Margadonna,Muro,Bobev}.
Moreover, a similar lattice expansion is seen for the other samples, $x$ = 0.9, 0.94, and 1.2,
where $a$, $c$, and the lattice volumes show a conventional decrease on cooling.
This behavior well agrees with the divalent Yb state in these systems,
because Yb$^{2+}$ ions can no longer expand through a valence transition mechanisms.

In summary, we have determined the solubility range of the YbGa$_x$Ge$_{2-x}$ system.
In the composition range of $0.9 \leq x \leq 1.5$,
it is found that single phase samples isostructural to YbGaGe are formed.
The magnetic susceptibility of these samples reveals that the Yb ions are consistently divalent.
This should lead a normal lattice shrinkage with decreasing temperature.
In fact, X-ray diffraction shows normal lattice shrinkage on cooling in these systems.
These results lead us the conclusion that the valence state of Yb in YbGa$_x$Ge$_{2-x}$ is not sensitive
to the composition $x$ but is robustly divalent, 
and that a zero thermal expansion does not occur in YbGa$_x$Ge$_{2-x}$
for whole the composition range ($0.9 \leq x \leq 1.5$),
at least not via a valence transition mechanism.

The YbGaGe phase with Yb$^{3+}$ ions may be obtained at much smaller $x$,
because Yb$^{3+}$ favors smaller lattice volumes.
However, such samples were not obtained by conventional arc melting or annealing technique.
Other method such as high pressure synthesis may be useful.

The authors sincerely thank T. Yokoya for fruitful discussions and comments.
They also thank T. Takabatake for sending us their preprint.

\end{document}